\documentclass[prl,twocolumn,superscriptaddress,showpacs]{revtex4-1}
\usepackage{graphics}
\usepackage{color}
\usepackage{amsmath}
\usepackage{amssymb}
\usepackage{hyperref}
\usepackage{graphicx}
\newcounter{fnnumber}

\begin{document}
\title{Quantum oscillation in narrow-gap topological insulators}
\author{Long Zhang}
\affiliation{International Center for Quantum Materials, School of Physics, Peking University, Beijing, 100871, China}
\author{Xue-Yang Song}
\affiliation{International Center for Quantum Materials, School of Physics, Peking University, Beijing, 100871, China}
\author{Fa Wang}
\affiliation{International Center for Quantum Materials, School of Physics, Peking University, Beijing, 100871, China}
\affiliation{Collaborative Innovation Center of Quantum Matter, Beijing, 100871, China}
\pacs{71.10.-w, 71.20.Eh, 71.28.+d}
\begin{abstract}

The canonical understanding of quantum oscillation in metals is challenged by the observation of de Haas-van Alphen effect in an insulator, SmB$_{6}$ [Tan \emph{et al}, Science {\bf349}, 287 (2015)]. Based on a two-band model with inverted band structure, we show that the periodically narrowing hybridization gap in magnetic fields can induce the oscillation of low-energy density of states in the bulk, which is observable provided that the activation energy is small and comparable to the Landau level spacing. Its temperature dependence strongly deviates from the Lifshitz-Kosevich theory. The nontrivial band topology manifests itself as a nonzero Berry phase in the oscillation pattern, which crosses over to a trivial Berry phase by increasing the temperature or the magnetic field. Further predictions to experiments are also proposed.

\end{abstract}
\date{\today}

\maketitle

\begin{figure}[!bt]
\centering
\includegraphics[width=0.48\textwidth]{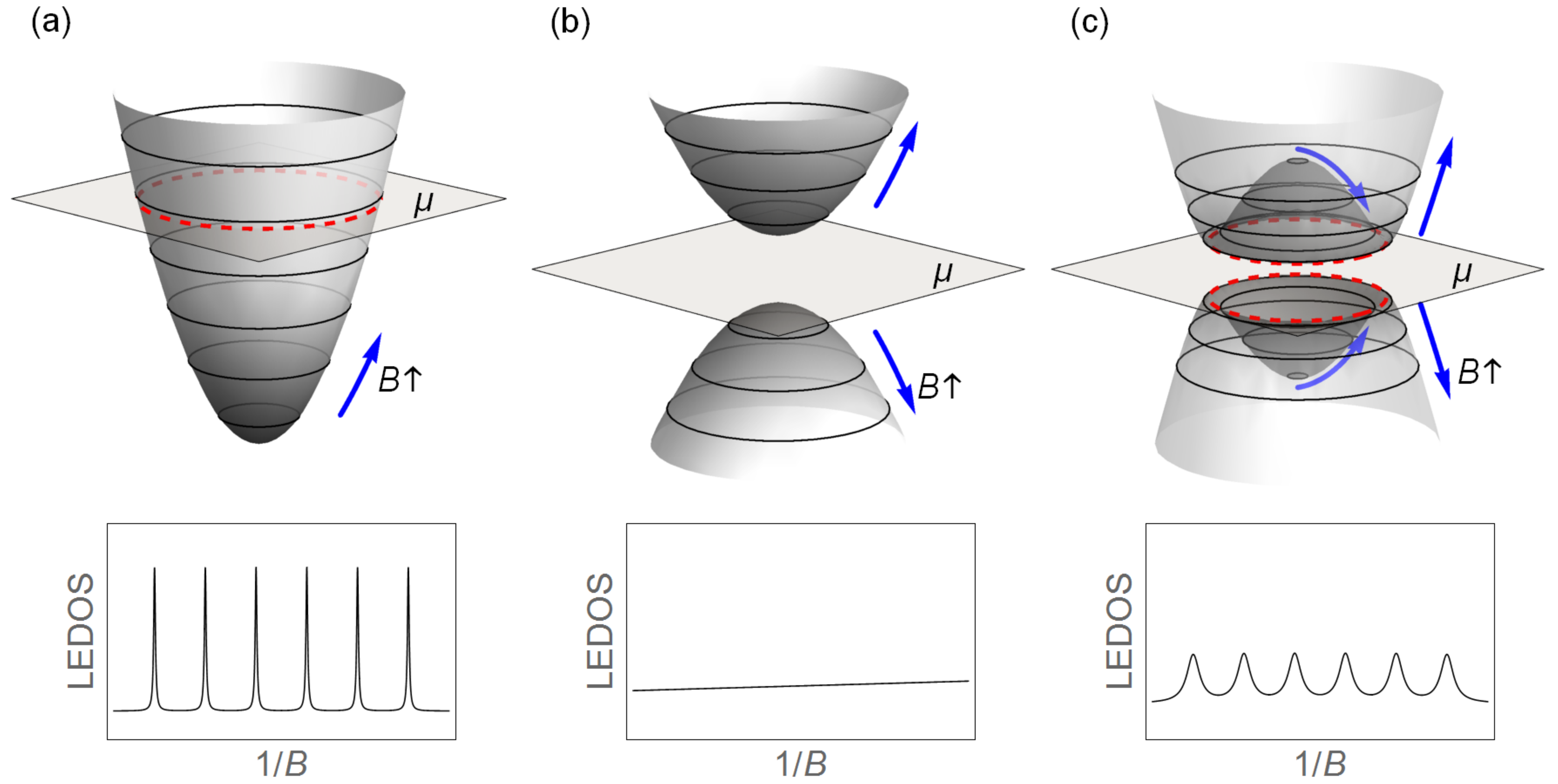}
\caption{(Color online) Illustration of LLs and possible low-energy DOS oscillation in (a) metals, (b) parabolic band insulators and (c) insulators with an inverted band structure. The planes and the thin circles denote the chemical potential and LL orbits respectively. The blue arrows indicate the flow of LLs with increasing magnetic fields. The red dashed circles denote the Fermi surface in (a) and the band edges in (c).}
\label{fig:intuitive}
\end{figure}

\emph{Introduction.---}Quantum oscillation is a nontrivial manifestation of Landau quantization in metals \cite{Shoenberg1984magnetic}. In a uniform magnetic field, an electron makes cyclotron motion with a conserved energy. If a constant energy surface forms a closed orbit in the reciprocal space, the quantization condition dictates that the area $A$ enclosed by the orbit satisfies,
\begin{equation} \label{eq:Bohr}
\frac{A\hbar}{e}\frac{1}{B}=2\pi (n+\gamma), \quad n\in \mathbb{N}.
\end{equation}
So the single-particle eigenstates form Landau levels (LLs). In metals, the chemical potential intersects an energy band, so the density of states (DOS) near the chemical potential peaks periodically as LLs cross the chemical potential with the variation of $1/B$ with the frequency given by $F=A_{F}\hbar/(2\pi e)$, in which $A_{F}$ is the area of the Fermi surface [see Fig. \ref{fig:intuitive} (a)] \footnote{This argument assumes a fixed chemical potential. In general, the chemical potential must shift to preserve the electron density, but its variation does not cancel the quantum oscillation. In particular, for the two-band model studied in this work, the chemical potential does not shift at all.}. The DOS oscillation results in the oscillation of various physical quantities, e.g, the magnetic susceptibility (de Haas-van Alphen effect) and the resistivity (Shubnikov-de Haas effect).

The constant $\gamma$ in Eq. (\ref{eq:Bohr}) is directly related to the Berry phase $\phi_{B}$ the electron accumulates during a cyclotron period, $2\pi(\gamma-1/2)=-\phi_{B}$ \cite{Mikitik1999}. $\gamma$ determines the positions of peaks and dips in the oscillation and can be extracted with the Landau level index analysis \cite{Novoselov2005, Zhang2005, Taskin2011, Li2013e}.

This canonical understanding of quantum oscillation is challenged by the recent observation of de Haas-van Alphen effect in an insulator, SmB$_{6}$ \cite{Tan2015}. SmB$_{6}$ has a narrow thermal activation gap in its bulk states, $\Delta \simeq 40 \mathrm{K}$ even in strong magnetic fields \cite{Cooley1995, Tan2015}. It is argued that the high-frequency quantum oscillation originates in the bulk states, as opposed to the topologically protected metallic surface states \cite{Li2013e} (however, cf. Ref. \cite{Erten2015} for a different interpretation), which is a consequence of the proposal of SmB$_{6}$ as a topological Kondo insulator \cite{Dzero2010, Dzero2012, Alexandrov2013, Lu2013a}. The temperature dependence deviates from the Lifshitz-Kosevich (LK) theory \cite{Tan2015}. So it is interesting to check the possibility of the insulating bulk states displaying quantum oscillation. Furthermore, it is desirable to find if any signature of the nontrivial band topology arises in the quantum oscillation.

Before proceeding to detailed model study, we first present an intuitive argument based on the semiclassical treatment of the Landau quantization as illustrated in Fig. \ref{fig:intuitive}. In contrast to the metals, in an insulator with parabolic bands either filled or empty, all LLs flow away from the chemical potential as the magnetic field increases, so the low-energy DOS [defined in Eq. (\ref{eq:LDoS})] decreases monotonically and does not oscillate at all. However, if the insulator has an inverted band structure as shown in Fig. \ref{fig:intuitive} (c), which is modelled by the two-band Hamiltonian in Eq. (\ref{eq:model}), as the magnetic field increases, LLs periodically approach the band edges, i.e., the bottom of the conduction band and the top of the valence band, resulting in periodic narrowing of the hybridization gap and low-energy DOS oscillation. Therefore, the band edges play a similar role as the Fermi surface in metals and the oscillation frequency is proportional to the enclosed area $A_{\mathrm{edge}}$,
\begin{equation} \label{eq:freq}
F=\frac{\hbar}{2\pi e}A_{\mathrm{edge}}.
\end{equation}
The oscillation is observable only if the amplitude of gap narrowing, which is related to the LL spacing near the band edge, is comparable to the activation gap itself. For a narrow hybridization gap, $A_{\mathrm{edge}}$ roughly equals the Fermi pocket area of the metal in the absence of hybridization. This semiclassical picture will be adopted again to show that there is a nontrivial Berry phase in the quantum oscillation pattern as a consequence of the nontrivial band topology. The temperature dependence of the oscillation amplitude is found to strongly deviate from the LK theory. Further predictions to experiments will also be discussed.


\emph{Model.---}We shall study the following two-band model in the continuum,
\begin{equation} \label{eq:model}
H=\sum_k \begin{pmatrix}
d_k^\dagger & f_k^\dagger
\end{pmatrix} 
\begin{pmatrix}
\frac{k^2}{2m_d}-\mu_d & V\vec{k}\cdot \vec{\sigma}\\
V\vec{k}\cdot \vec{\sigma} & -\frac{k^2}{2m_f}-\mu_f
\end{pmatrix}
\begin{pmatrix}
d_k\\
f_k
\end{pmatrix},
\end{equation}
in which $d_{k}=(d_{k\uparrow},d_{k\downarrow})^{T}$ and $f_{k}=(f_{k\uparrow},f_{k\downarrow})^{T}$ are $d$- and $f$-band electrons with pseudospin-$1/2$. $\vec{\sigma}$ are the Pauli matrices acting on the pseudospin space. If $\delta \mu\equiv \mu _{d}-\mu _{f}>0$, the model has an inverted band structure. The electron-like $d$-band and the hole-like $f$-band are hybridized by the parity-odd $V\vec{k}\cdot\vec{\sigma}$ term and open a finite gap. If the chemical potential lies within the gap, this model describes topological insulators in 2D and 3D \cite{Bernevig2006, Bernevig2006a, Fu2007a, Fu2007, Hasan2010, Qi2011}.

There are four bands in SmB$_{6}$ with pseudospin-$1/2$ near the chemical potential and the band inversion happens around the three $X$ points \cite{Lu2013a, Alexandrov2013, Yu2015}. Eq. (\ref{eq:model}) can be taken as a simplified two-band $k\cdot p$ model expanded around one $X$ point \cite{Yu2015, Alexandrov2015}. We adopt the following band parameters derived from a tight-binding model \footnote{See Supplemental Material [url], which includes Refs \cite{Allais2014, Zhang2014b, Zhang2014c}.}\setcounter{fnnumber}{\thefootnote} throughout this work unless specified otherwise, $m_{d}=\hbar^{2}/2ta_{0}^{2}$, $\alpha\equiv m_{d}/m_{f}=0.1$, $\delta \mu=0.5t$. The $d$-band hopping amplitude $t$ is set to be unity. A weak hybridization $V/a_{0}=0.015t$ leads to a narrow gap $\Delta_{g}= 0.012t$. Substituting $t\simeq 640\mathrm{meV}$ estimated from the calculated SmB$_{6}$ band structure \cite{Yu2015}, one finds $\Delta_{g}=7.7\mathrm{meV}$, which roughly equals two times of the $40\mathrm{K}$ activation energy. Therefore, our model captures the main features of the SmB$_{6}$ band structure. The strongest magnetic field in experiments $\sim 50\mathrm{T}$ corresponding to $1/500$ flux quanta per unit cell is covered in our calculations. The Zeemann effect estimated in experiments is quite weak \cite{Cooley1995, Tan2015} and does not qualitatively change our results, so will be neglected in our presentation.

The possible quantum oscillation \emph{from the bulk states} is characterized by the low-energy DOS (LEDOS) \emph{near} the chemical potential, defined as the broadened DOS at temperature $T$,
\begin{equation} \label{eq:LDoS}
D_{T}=\int_{-\infty}^{+\infty}d\xi \frac{\partial n_{F}(\xi-\mu,T)}{\partial \mu}D(\xi)=\sum_{i}\frac{\partial n_{F}(\epsilon_{i}-\mu,T)}{\partial \mu},
\end{equation}
in which $D(\xi)$ is the single-particle DOS. The summation on the right hand of Eq. (\ref{eq:LDoS}) is taken over the single-particle energy spectrum. LEDOS is related to various physical quantities at finite temperature, e.g., the Pauli susceptibility, the compressibility and the resistivity, and its oscillation necessarily results in the oscillation of these quantities. Besides, an advantage in calculating LEDOS is that it does not require any regularization procedure. In contrast, the free energy is (formally) divergent due to the hole-like $f$-band. Upon regularization, a cutoff at some negative energy may play a similar role as the Fermi surface in metals and result in artificial oscillation, which is avoided in the LEDOS calculations.


\begin{figure}[!bt]
\centering
\includegraphics[width=0.48\textwidth]{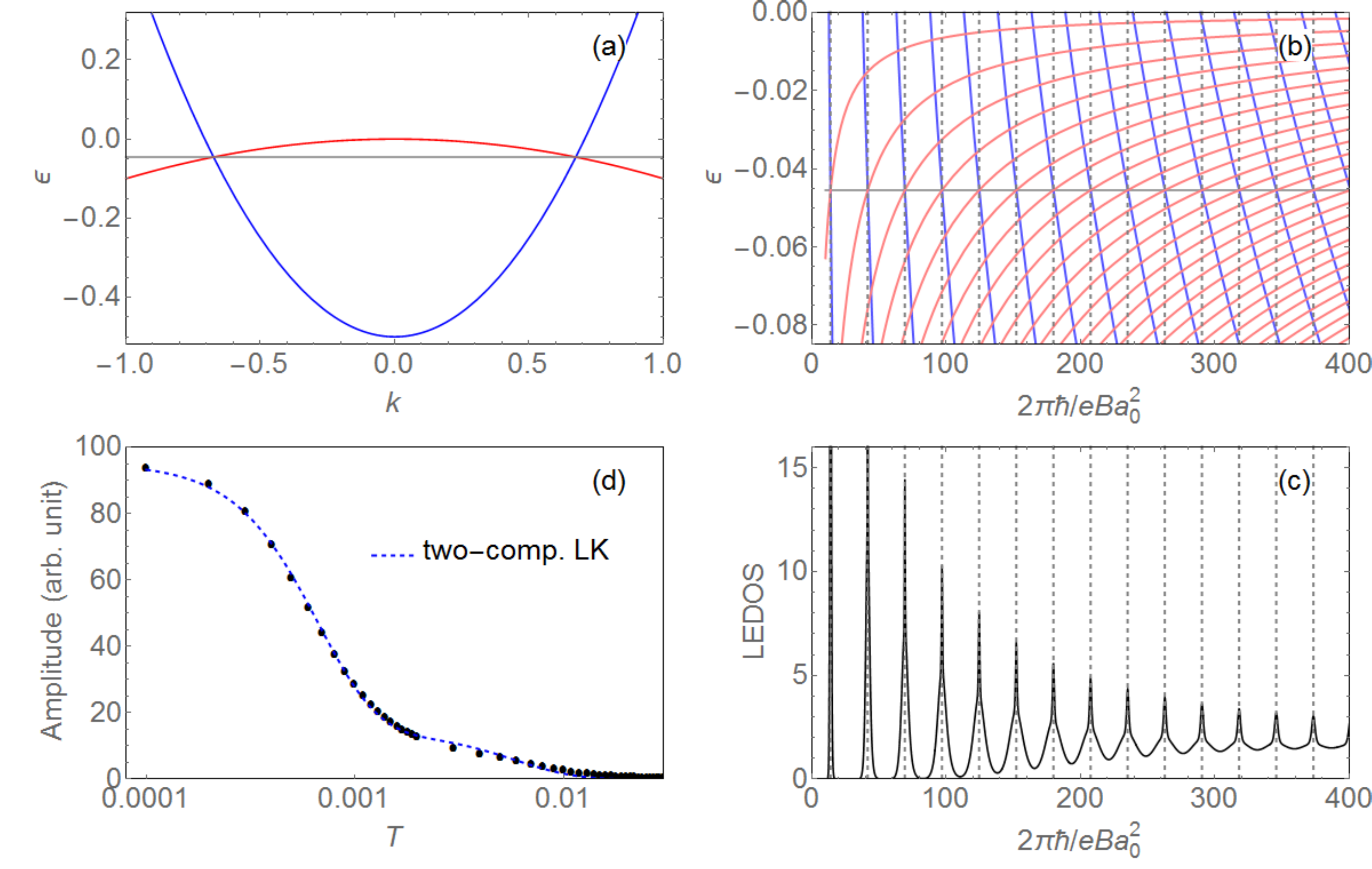}
\caption{(Color online) (a) band structure, (b) LL spectrum and (c) LEDOS oscillation of the semimetal in 2D. The dashed gray lines in (b) and (c) indicate whenever Eq. (\ref{eq:Bohr}) is satisfied at the chemical potential with $\gamma=1/2$. (d) The temperature dependence of the oscillation amplitude (black dots) and the fitting by the two-component LK formula (dashed blue curve). Band structure parameters are specified in the main text with $t$ set to be unity. The abscissas in (b) and (c) are labelled with the inverse number of flux quanta per unit cell.}
\label{fig:semimetal}
\end{figure}

\emph{2D semimetal.---}If the hybridization is turned off, $V=0$, the chemical potential lies exactly where the $d$- and $f$-bands intersect, forming an electron-like and a hole-like Fermi pockets with equal size [Fig. \ref{fig:semimetal} (a)]. In magnetic fields, these bands form two sets of LLs,
\begin{equation}
\epsilon_{n}^{d}=\frac{eB\hbar}{m_{d}}\bigg(n+\frac{1}{2}\bigg)-\mu _{d},\quad \epsilon_{n}^{f}=-\frac{eB\hbar}{m_{f}}\bigg(n+\frac{1}{2}\bigg)-\mu _{f},\quad n\in \mathbb{N}.
\end{equation}
LLs cross the Fermi surface periodically and result in the LEDOS oscillation as shown in Figs. \ref{fig:semimetal} (b) and (c).

The temperature dependence of the oscillation amplitude has an unusual two-plateau feature [Fig. \ref{fig:semimetal} (d)], which resembles that found in SmB$_{6}$ \cite{Tan2015}. The reason is that both Fermi pockets contribute to the LEDOS oscillation with equal frequency. At finite temperature, the contribution from each band is captured by the LK theory, so the total oscillation amplitude is described by the \emph{two-component} LK formula,
\begin{equation} \label{eq:doubleLK}
R_{T}=c_{d}\frac{\chi_{d}}{\sinh \chi_{d}}+c_{f}\frac{\chi_{f}}{\sinh\chi_{f}},
\end{equation}
in which $\chi_{d,f}=2\pi^{2}m_{d,f}T/eB\hbar$, $c_{d,f}\propto m_{d,f}$. The oscillation amplitudes are extracted as the heights of the dominant Fourier peaks, which are fitted perfectly by Eq. (\ref{eq:doubleLK}) \footnotemark[\thefnnumber].


\emph{2D topological insulator.---}In a magnetic field, the $V\vec{k}\cdot \vec{\sigma}$ term is replaced by $V(\vec{k}-e\vec{A}/\hbar)\cdot \vec{\sigma}$, which hybridizes different LLs. In 2D, the Hamiltonian is decoupled into two sectors, the $d_{\uparrow}$-$f_{\downarrow}$ ($\uparrow\downarrow$) sector and the $d_{\downarrow}$-$f_{\uparrow}$ ($\downarrow\uparrow$) sector. Within each sector, the LLs are hybridized \emph{obliquely}, i.e., the $n$th $d_{\uparrow}$-LL is hybridized with the $(n-1)$th $f_{\downarrow}$-LL, while the $(n-1)$th $d_{\downarrow}$-LL with the $n$th $f_{\uparrow}$-LL, forming the following spectrum,
\begin{align}
\epsilon_{n\pm}^{\uparrow\downarrow}&=\frac{1}{2}\Bigg(\epsilon_{n}^{d}+\epsilon_{n-1}^{f}\pm\sqrt{(\epsilon_{n}^{d}-\epsilon_{n-1}^{f})^{2}+8nV^{2}eB/\hbar}\Bigg),\\
\epsilon_{n\pm}^{\downarrow\uparrow}&=\frac{1}{2}\Bigg(\epsilon_{n-1}^{d}+\epsilon_{n}^{f}\pm\sqrt{(\epsilon_{n-1}^{d}-\epsilon_{n}^{f})^{2}+8nV^{2}eB/\hbar}\Bigg),
\end{align}
for $n\geq 1$. The $d_{\uparrow}$- and $f_{\uparrow}$-LLs with index $n=0$ are unaffected.

\begin{figure}[!bt]
\centering
\includegraphics[width=0.48\textwidth]{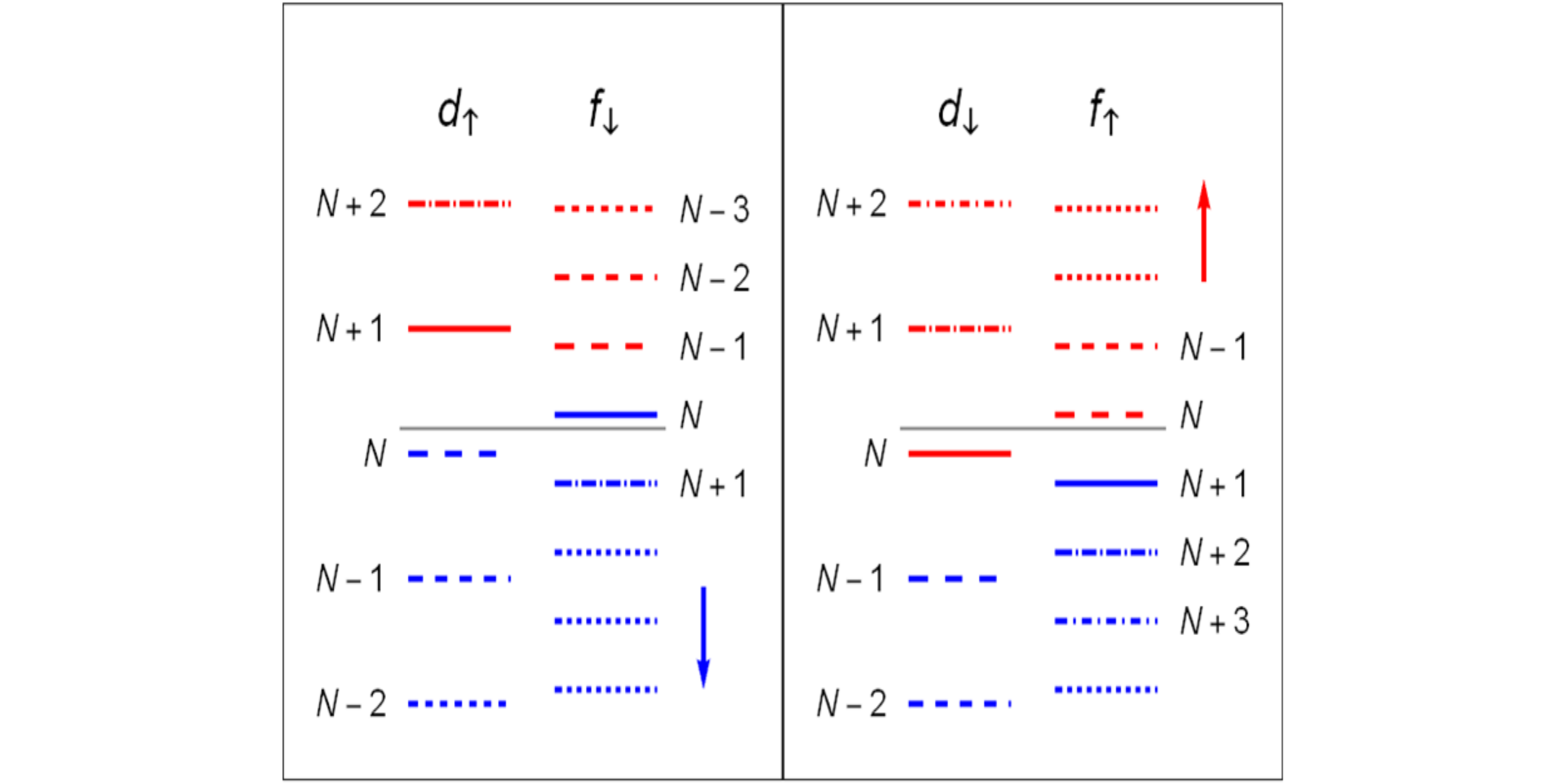}
\caption{(Color online) Illustration of the LL hybridization in 2D. In each sector, each $d$-LL hybridizes with an $f$-LL drawn in the same dashing style. LLs in red are pushed upward while the blue downward. The long gray bars denote the chemical potential. All LLs shift away from it except that in the $\{N,N+1\}$ LL pairs (solid bars), one LL out of each pair shifts toward it and may pass each other.}
\label{fig:levelrepulsion}
\end{figure}

Let us start from the semimetal without hybridization and dub the highest occupied $d$-LL index $N$, $N=\lfloor\delta\mu/\hbar\omega_{c}^{*}-1/2\rfloor$, with $\omega_{c}^{*}\equiv eB\hbar(m_{d}+m_{f})/m_{d}m_{f}$. The highest unoccupied $f$-LL index is also $N$. As the hybridization is turned on, all LLs are pushed away from the chemical potential, except one pair in each sector, the $(N+1)$th $d_{\uparrow}$-LL and the $N$th $f_{\downarrow}$-LL, and the $N$th $d_{\downarrow}$-LL and the $(N+1)$th $f_{\uparrow}$-LL, as illustrated in Fig. \ref{fig:levelrepulsion}. One LL out of each pair is pushed \emph{toward} the chemical potential. In weak magnetic fields, the level repulsion overcomes the small LL spacing and these two LLs pass each other, leaving a hybridization gap. If the hybridization is perturbatively small in strong magnetic fields, the LLs do not pass each other, so the spectrum near the chemical potential is largely unaffected and remains metallic. Therefore, the magnetic field induces a gap-closing transition from a topological insulator to a metal \footnotemark[\thefnnumber].

\begin{figure}[!bt]
\centering
\includegraphics[width=0.48\textwidth]{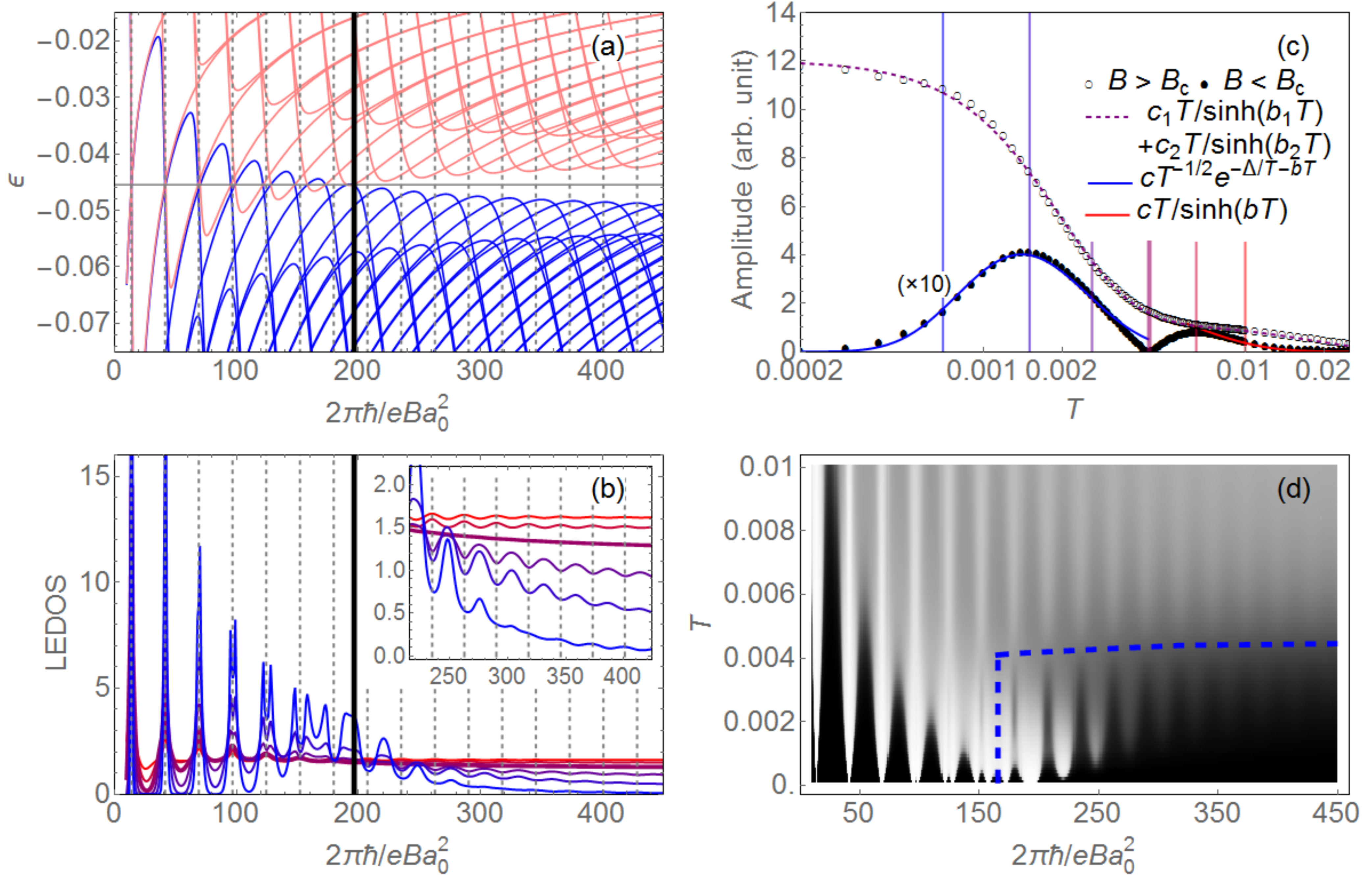}
\caption{(Color online) (a) LL spectrum and (b) LEDOS oscillation of the 2D model with a hybridization gap. The thick vertical lines indicate $B_{c}$, the critical field of gap-closing. The Inset of (b) highlights the phase jump at finite temperature. The LEDOS are plotted at temperatures indicated by the vertical lines in (c). (c) Temperature dependence of the oscillation amplitudes extracted in $B>B_{c}$ (open circles) and $B<B_{c}$ (filled circles, multiplied by 10 for clarity). (d) Intensity plot of LEDOS, highlighting the $\pi$ phase jump around $B\simeq B_{c}$ and $T\simeq \Delta$, which are indicated by the dashed curves.}
\label{fig:2dhybridized}
\end{figure}

The energy spectrum in magnetic fields is plotted in Fig. \ref{fig:2dhybridized} (a). The hybridization gap is closed above a critical field $B_{c}$. For $B>B_{c}$, the low energy spectrum is nearly the same as the unhybridized case, resulting in similar LEDOS oscillation. The temperature dependence is captured by the two-component LK formula, as shown in Fig. \ref{fig:2dhybridized} (c).

For $B<B_{c}$, a close inspection on the LL spectrum finds periodic gap narrowing as expected from the semiclassical argument, which leads to the smooth oscillation of LEDOS. However, we find various peculiarities detailed below.

The oscillation amplitude has a non-monotonic temperature dependence in sharp contrast to the LK theory. At low temperature, the amplitude has a broad hump, which is a consequence of the activation gap $\Delta$. For $T\ll \Delta$, the oscillation amplitude is captured by the asymptotic formula \footnotemark[\thefnnumber],
\begin{equation} \label{eq:asymp}
R_{T}\sim T^{-1/2}e^{-\Delta/T-bT},
\end{equation}
in which the $e^{-\Delta/T}$ factor comes from the thermal activation while the $e^{-bT}$ factor reflects the thermal smearing similar to the LK theory.

Around $T\simeq \Delta$, the amplitude dips to zero, which coincides with a $\pi$ phase jump in the oscillation pattern, as highlighted in the Inset of Fig. \ref{fig:2dhybridized} (b). Furthermore, at low temperature, there is another (approximate to) $\pi$ phase jump around $B_{c}$, while the high-temperature oscillation varies smoothly across $B_{c}$ without any phase change as shown in Fig. \ref{fig:2dhybridized} (d). Therefore, the Berry phase changes by $\pi$ in both circumstances.

The phase jump across $B_{c}$ is actually implied by the LL spectrum, in which the periodic gap narrowing for $B<B_{c}$ is replaced by periodic widening for $B>B_{c}$ due to the gap closing, and the LEDOS peaks are replaced by dips correspondingly. Moreover, the nonzero Berry phase for $B<B_{c}$ turns out to be a manifestation of the oblique hybridization. Let us turn back to the semiclassical picture and focus on one LL in the $d_{\downarrow}$-$f_{\uparrow}$ sector with energy $\epsilon_{n+}^{\downarrow\uparrow}$, which comes from the hybridization between the $(n-1)$th $d_{\downarrow}$-LL and the $n$th $f_{\uparrow}$-LL. As it flows from the $f$-band top downward to the conduction band bottom and upward again along the $d$-band with increasing $B$, the total phase in Eq. (\ref{eq:Bohr}) changes by $-2\pi$, implying that the Berry phase near the band bottom is approximate to $\pi$, consistent with the LEDOS oscillation pattern at low temperature. At high temperature, $T>\Delta$, LLs with trivial Berry phase dominate over those near the band edges, leading to the crossover of the oscillation phase.

\begin{figure}[!bt]
\centering
\includegraphics[width=0.48\textwidth]{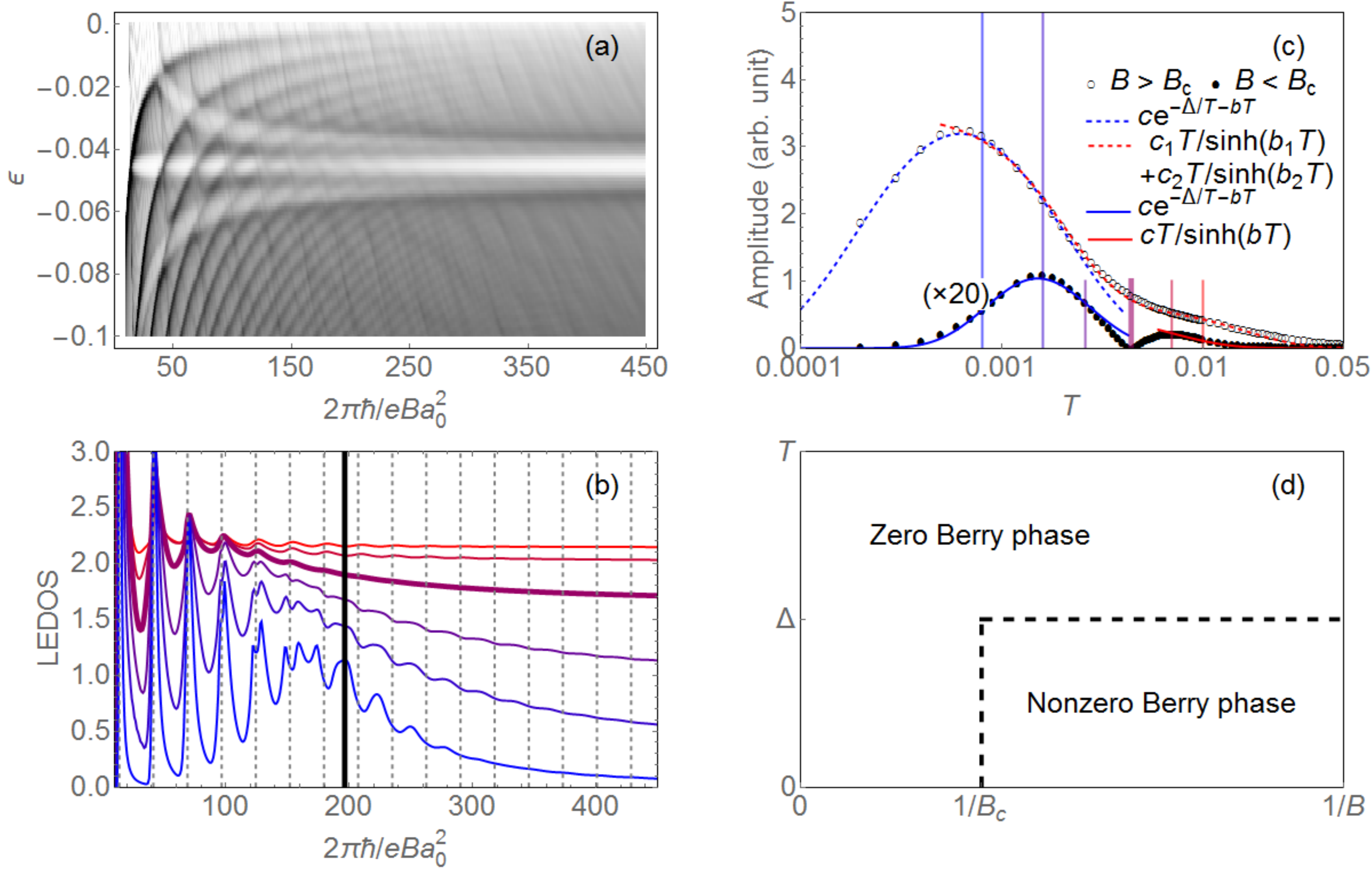}
\caption{(Color online) (a) single-particle DOS of the 3D model and (b) LEDOS oscillation. The thick vertical line indicates $B_{c}$. The LEDOS are plotted at temperatures indicated by the vertical lines in (c). (c) Temperature dependence of the oscillation amplitudes for $B>B_{c}$ (open circles) and $B<B_{c}$ (filled circles, multiplied by 20 for clarity). (d) Schematic illustration of Berry phases in different regimes.}
\label{fig:3D}
\end{figure}

\emph{Quantum oscillation in 3D.---}The two-band model Eq. (\ref{eq:model}) is easily generalized to 3D with the $z$-components included in $\vec{k}$ and $\vec{\sigma}$. The LEDOS oscillation shown in Fig. \ref{fig:3D} is qualitatively similar to the 2D case with minor modifications.

First, in the oscillation frequency formula Eq. (\ref{eq:freq}), $A_{\mathrm{edge}}$ should be understood as the area enclosed by the extremum orbit on the band edges. For weak hybridization, it roughly equals that of the Fermi surface in the absence of hybridization.

Second, the $k_{z}\sigma_{z}$ term introduces further hybridization between the $n$th $d_{\uparrow}$ ($d_{\downarrow}$)-LL and the $n$th $f_{\uparrow}$ ($f_{\downarrow}$)-LL and opens a gap in the $B>B_{c}$ regime for nonzero $k_{z}$ ($B_{c}$ is defined as the gap-closing field for $k_{z}=0$), resulting in a persistent gap in DOS as shown in Fig. \ref{fig:3D} (a). As a result, the LEDOS oscillation in the $B>B_{c}$ regime also shows thermal activation behavior at low temperature, which is captured by an asymptotic formula similar to Eq. (\ref{eq:asymp}), $R_{T}\sim e^{-\Delta'/T-bT}$. Due to the gap nodes at $k_{z}=0$ at particular field strengths, $\Delta'$ is much smaller than the gap away from these fields. So the temperature with the maximum oscillation amplitude $T_{\max}$ should be much lower than the activation energy measured with resistivity.

Otherwise the LEDOS oscillation in 3D carries all essential features as in the 2D case. For $B>B_{c}$, the temperature dependence is captured by the two-component LK formula for $T> T_{\max}$. For $B<B_{c}$, the nontrivial Berry phase shows up, which crosses over to the trivial Berry phase at high temperature or high magnetic fields.


\emph{Summary and discussion.---}To summarize, we find that an insulator with inverted bands can show quantum oscillation in its bulk low-energy DOS due to the periodic gap-narrowing in magnetic fields. The oscillation frequency is proportional to the area enclosed by the extremum orbit on the band edge. For a topological insulator, the nontrivial band topology manifests itself as a nonzero Berry phase in the oscillation. The temperature dependence deviates from the LK theory and shows thermal activation behavior at low temperature in particular. These features are also reproduced by a tight-binding model on the lattice \footnotemark[\thefnnumber].

In a recent publication \cite{Knolle2015}, the authors found quantum oscillation in a similar two-band model. The oscillation frequency is consistent with our result. However, the hybridization term in their work is parity-even, so the hybridization gap is topologically trivial. The nonzero Berry phase and the bulk gap-closing at $B_{c}$ found in our work are missing.

Several features can be tested in experiments. First is the sizeable periodic gap-narrowing in magnetic fields that causes the LEDOS oscillation, which can be extracted from the resistivity or with infrared spectroscopy. Second is the thermal activation behavior, i.e., the decreasing oscillation amplitude at temperature much lower than the activation energy. Third is the nonzero Berry phase. Even if it is difficult to extract the Berry phase directly \cite{Taskin2011}, it is possible to observe a $\pi$ phase jump at the boundaries sketched in Fig. \ref{fig:3D} (d).

\emph{Note added.---}Upon completion of this work, we became aware of Ref. \cite{Erten2015}, in which a different scenario for the quantum oscillation in SmB$_{6}$ was suggested.


L.Z. is grateful to Z. Fang, X.-J. Liu, Z.-Y. Meng and R. Yu, and in particular to X. Dai, S.-K. Jian and D.-H. Lee for stimulating discussions and valuable suggestions, and to J.-W. Mei for previous collaborations on related topics. This work was supported by the National Key Basic Research Program of China (Grant No. 2014CB920902) and the National Science Foundation of China (Grant No. 11374018).

\bibliography{/Dropbox/ResearchNotes/BibTex/library}

\end{document}


\title{Supplemental materials for \emph{Quantum oscillation in narrow-gap topological insulators}}
\author{Long Zhang}
\affiliation{International Center for Quantum Materials, School of Physics, Peking University, Beijing, 100871, China}
\author{Xue-Yang Song}
\affiliation{International Center for Quantum Materials, School of Physics, Peking University, Beijing, 100871, China}
\author{Fa Wang}
\affiliation{International Center for Quantum Materials, School of Physics, Peking University, Beijing, 100871, China}
\affiliation{Collaborative Innovation Center of Quantum Matter, Beijing, 100871, China}

\date{\today}

\maketitle

\section{Two-component Lifshitz-Kosevich formula}

In a 2D metal with a parabolic band, the LEDOS is given by
\begin{equation}
D_{T}=\sum_{i}\frac{\partial n_{F}(\epsilon_{i}-\mu,T)}{\partial \mu}=\frac{eB}{2\pi \hbar}\frac{1}{2T}\sum\limits_{n=0}^{\infty}\frac{1}{\cosh(\epsilon_{n}-\mu)/T+1},
\end{equation}
in which $\epsilon_{n}=(eB\hbar/m)(n+1/2)$ is the LL spectrum. The prefactor $eB/2\pi\hbar$ is the LL degeneracy in a unit system size. Using the Poisson resummation formula, the oscillatory component is given by
\begin{equation} \label{eq:twoLK}
\begin{split}
D_{T}&=\frac{eB}{2\pi\hbar}\frac{1}{2T}\sum\limits_{k=1}^{\infty}\int_{-\infty}^{+\infty}dn\frac{1}{\cosh(\epsilon_{n}-\mu)/T+1}e^{2\pi i k n}+\cdots\\
&=\frac{eB}{2\pi \hbar}\frac{m}{2eB\hbar}\sum\limits_{k=1}^{\infty}\frac{2\pi^{2}mkT/eB\hbar}{\sinh(2\pi^{2}mkT/eB\hbar)}e^{2\pi ik(\mu m/eB-1/2)}+\cdots,
\end{split}
\end{equation}
in which the ellipses denote non-oscillatory component. Except for the well-known LK reduction factor $\chi/\sinh \chi$ with $\chi=2\pi^{2}mkT/eB\hbar$, there is a prefactor proportional to $m$ in the LEDOS oscillation amplitude, which shows up if we add the contributions from more than one Fermi pocket with equal area. The pocket with a heavier effective mass contributes a larger oscillation amplitude in the zero-temperature limit in particular.

In the main text, in fitting Eq. (6) to the oscillation amplitudes of the half metal, we leave $\chi_{d,f}/T$ and $c_{d,f}$ as free parameters for best fitting. We indeed find that in the zero-temperature limit, the $f$-band plateau is higher than the $d$-band plateau. For reference, the best-fitting parameters in Fig. 2 (d) are $c_{d}=13.99\pm 0.37$, $c_{f}=81.08\pm 0.52$, $\chi_{d}/T=428\pm 17$, $\chi_{f}/T=3775\pm 37$, so we find $\chi_{d}/\chi_{f}=0.11$, $c_{d}/c_{f}=0.17$. Their difference from the mass ratio $m_{d}/m_{f}=0.1$ may be attributed to the magnetic field strength $B$ dependence in Eq. (\ref{eq:twoLK}). Because the Fourier transformation is taken over a broad range of $B$ where the oscillation amplitude varies wildly, the effect of averaging over $B$ is different on the $f$ and the $d$ components.

The LEDOS is proportional to the Pauli paramagnetic susceptibility at finite temperature. In contrast, the orbital diamagnetization can be deduced from the free energy. The oscillation amplitude of the orbital diamagnetization has a prefactor proportional to $1/m$ besides the LK factor $\chi/\sinh \chi$ \cite{Shoenberg1984magnetic}, so the Fermi pocket with a heavier effective mass contributes smaller, which is different from the Pauli susceptibility.

In a real sample of half metal, both Pauli and orbital magnetization contribute, so the relative height of the $d$- and $f$-band plateaus depends on which contribution dominates. We may suggest the following thumb rule. If the sample is paramagnetic, the Pauli susceptibility dominates and the $f$-band plateau in the zero-temperature limit should be higher, otherwise the $d$-band plateau should be higher.

\section{Magnetic field induced gap-closing}

\begin{figure}
\centering
\includegraphics[width=0.6\textwidth]{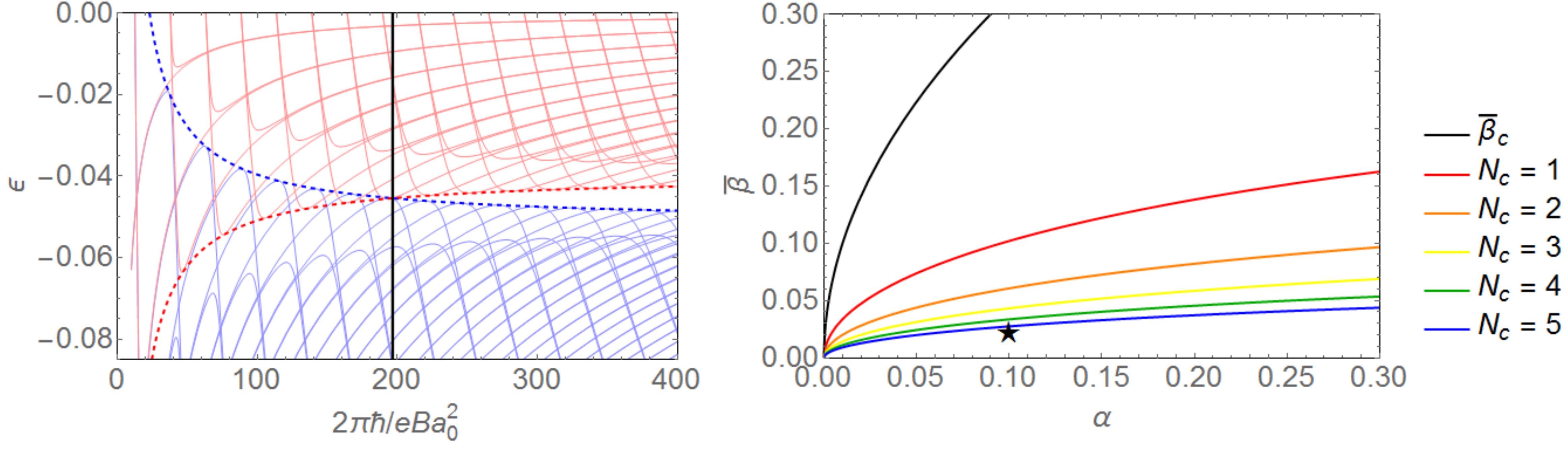}
\caption{Left: the LL spectrum in magnetic fields with the band edges characterized by the enveloping curves (blue dashed: $\epsilon_{\mathrm{env}-}^{\uparrow\downarrow}$, red dashed: $\epsilon_{\mathrm{env}+}^{\downarrow\uparrow}$). The vertical line indicates the critical field strength $B_{c}$ where the envelopes intersect with each other. Band parameters are the same as those adopted in the main text. Right: phase diagram illustrating the possibility of a gap-closing transition in magnetic fields and the number of oscillation periods that can be observed in the $b>b_{c}$ regime. The star indicates the parameters adopted in the main text.}
\label{fig:bc}
\end{figure}

In the 2D model, the magnetic field can induce a gap-closing transition from a topological insulator to a metal. As shown in Fig. \ref{fig:bc} (a), although the band edges vary irregularly with the magnetic field, they can be characterized by smooth envelopes. The envelope of a family of curves given by $F(x,y;n)=0$ ($n$ labels different curves) is the solution of the following equations,
\begin{equation}
\begin{cases}
F(x,y;n)=0;\\
\partial F(x,y;n)/\partial n=0.
\end{cases}
\end{equation}
To simplify notations, we introduce the dimensionless parameters $\alpha$, $b$ and $\beta$: $\alpha\equiv m_{d}/m_{f}$, $b\equiv eBa_{0}^{2}/\hbar$, $\beta\equiv V/a_{0}t$ and set $t$ to be unity. The envelope of the $\epsilon_{n-}^{\uparrow\downarrow}$ band is given by
\begin{equation}
\epsilon_{\mathrm{env}-}^{\uparrow\downarrow}=(1+\alpha)^{-2}\Bigg(\alpha(1+\alpha)(2b-\delta \mu)-(1-\alpha)\beta^{2}-2\beta \alpha^{1/2}\sqrt{(1+\alpha)\delta \mu -\beta^{2}-(1-\alpha^{2})b}\Bigg)-\mu _{f},
\end{equation}
which characterizes the top of the valence band in magnetic fields. For the $\epsilon_{n+}^{\downarrow\uparrow}$ band, if $\beta^{2}<\alpha\delta \mu$, the envelope is given by
\begin{equation}
\epsilon_{\mathrm{env}+}^{\downarrow\uparrow}=(1+\alpha)^{-2}\Bigg(-\alpha(1+\alpha)(2b+\delta\mu)-(1-\alpha)\beta^{2}+2\beta\alpha^{1/2}\sqrt{(1+\alpha)\delta\mu-\beta^{2}+(1-\alpha^{2})b}\Bigg)-\mu _{f};
\end{equation}
otherwise its lower edge is simply $\epsilon_{0}^{f}=-\alpha b-\mu _{f}$. As $b\rightarrow 0$,
\begin{align}
\epsilon_{\mathrm{env}-}^{\uparrow\downarrow}(b=0)&=(1+\alpha)^{-2}\Bigg(-\alpha(1+\alpha)\delta\mu-(1-\alpha)\beta^{2}-2\beta\alpha^{1/2}\sqrt{(1+\alpha)\delta\mu-\beta^{2}}\Bigg)-\mu _{f};\\
\epsilon_{\mathrm{edge}+}^{\downarrow\uparrow}(b=0)&=
\begin{cases}
(1+\alpha)^{-2}\Bigg(-\alpha(1+\alpha)\delta\mu-(1-\alpha)\beta^{2}+2\beta\alpha^{1/2}\sqrt{(1+\alpha)\delta\mu-\beta^{2}}\Bigg)-\mu _{f}, &\beta^{2}<\alpha\delta\mu;\\
-\mu _{f},&\beta^{2}>\alpha\delta\mu.
\end{cases}
\end{align}
These are exactly the band edges in the absence of magnetic fields.

If $\epsilon_{\mathrm{env}+}^{\uparrow\downarrow}(b)$ and $\epsilon_{\mathrm{edge}-}^{\downarrow\uparrow}(b)$ intersect with each other at $b_{c}>0$, the hybridization gap closes for $b>b_{c}$, as shown in Fig. \ref{fig:bc} (a). For $\beta^{2}/\delta\mu<\alpha$, i.e., the hybridization is relatively weak, $b_{c}$ is given by the solution of the following equation,
\begin{equation} \label{eq:bc}
2\alpha(1+\alpha)\bar{b}=\bar{\beta}\alpha^{1/2}\Bigg(\sqrt{1+\alpha-\bar{\beta}^{2}+(1-\alpha^{2})\bar{b}}+\sqrt{1+\alpha-\bar{\beta}^{2}-(1-\alpha^{2})\bar{b}}\Bigg),
\end{equation}
in which $\bar{\beta}^{2}\equiv \beta^{2}/\delta\mu$, $\bar{b}\equiv b/\delta \mu$. Eq. (\ref{eq:bc}) has a solution only if
\begin{equation}
\bar{\beta}^{2}<\min\bigg\{\frac{4\alpha}{1+\alpha},\alpha\bigg\}=\alpha\equiv \bar{\beta}_{c}^{2},\quad\mathrm{for}~\alpha<1,
\end{equation}
with the critical field strength $\bar{b}_{c}$ given by
\begin{equation}
\bar{b}_{c}=\frac{\bar{\beta}}{2\alpha}\sqrt{\frac{4\alpha}{1+\alpha}-\bar{\beta}^{2}}.
\end{equation}
For relatively strong hybridization $\beta^{2}/\delta\mu >\alpha$, the gap is not closed until the magnetic field is pushed to the quantum limit, i.e., the $n=0$ $f$-LL is pulled down below the top of the valence band. Therefore, the possibility of a magnetic field induced gap-closing depends on the band parameters $\alpha$, $\delta \mu$ and the hybridization $\beta$, as shown in Fig. \ref{fig:bc} (b). Along with it we also plot several contours with different $N_{c}\equiv N(b_{c})$, i.e., the highest occupied $d$-LL index at the gap-closing transition, which indicates the number of quantum periods that can be observed (in principle) in the $b>b_{c}$ regime.

\section{Oscillation amplitudes at low temperature}

Using the Poisson resummation formula, the LEDOS Eq. (4) in the main text in the insulating regime is cast into the following form,
\begin{equation} \label{eq:DT}
\begin{split}
D_{T}&=\frac{b}{2\pi}\int_{-\infty}^{\infty}d\xi\frac{\partial n_{F}(\xi-\mu,T)}{\partial\mu}\Bigg(\sum\limits_{n=1}^{\infty}\sum\limits_{\pm}\delta(\xi-\epsilon_{n\pm}^{\pm})+\delta(\xi-\epsilon_{0}^{d})+\delta(\xi-\epsilon_{0}^{f})\Bigg)\\
&=\frac{b}{\sqrt{2\pi}}\sum\limits_{k=1}^{\infty}\int_{0}^{\infty}dn\cos(2\pi kn)\int_{-\infty}^{+\infty}d\xi\frac{\partial n_{F}(\xi-\mu,T)}{\partial\mu}\sum\limits_{\pm}\delta(\xi-\epsilon_{n\pm}^{\pm})+\cdots\\
&=\frac{b}{\sqrt{2\pi}}\frac{1}{2T}\mathrm{Re}\sum\limits_{k=1}^{\infty}\int_{0}^{\infty}dne^{2\pi kin}\sum\limits_{\pm}\frac{1}{\cosh (\epsilon_{n\pm}^{\pm}-\mu)/T+1}+\cdots,
\end{split}
\end{equation}
in which $\epsilon_{n\pm}^{+}\equiv \epsilon_{n\pm}^{\uparrow\downarrow}$, $\epsilon_{n\pm}^{-}\equiv \epsilon_{n\pm}^{\downarrow\uparrow}$. The summation $\sum_{\pm}$ runs over all four sets of LLs. The ellipses denote non-oscillatory contributions. In the insulating regime, $|\epsilon_{n\pm}^{\pm}-\mu|\geq \Delta_{\pm}^{\pm}>0$ for all $n$. Expand $|\epsilon_{n\pm}^{\pm}-\mu|$ as a function of $n$ around the minima at $n_{\mathrm{gap}\pm}^{\pm}$,
\begin{equation} \label{eq:expansion}
|\epsilon_{n\pm}^{\pm}-\mu|\simeq \Delta_{\pm}^{\pm}+\frac{1}{2}|\epsilon_{n\pm}^{\pm\prime\prime}(n_{\mathrm{gap}\pm}^{\pm})|(n-n_{\mathrm{gap}\pm}^{\pm})^{2}.
\end{equation}
For a weak hybridization, $n_{\mathrm{gap}\pm}^{\pm}$ is approximate to the highest occupied $d$-LL index, $N=\lfloor\delta\mu/\hbar \omega_{c}^{*}-1/2\rfloor$, with $\omega_{c}^{*}\equiv \hbar eB(m_{d}+m_{f})/m_{d}m_{f}$. So we have
\begin{equation} \label{eq:cosh}
\frac{1}{\cosh (\epsilon_{n\pm}^{\pm}-\mu)/T+1}\simeq 2e^{-(\Delta_{\pm}^{\pm}+\frac{1}{2}|\epsilon_{n\pm}^{\pm\prime\prime}|(n-N)^{2})/T}.
\end{equation}
Substituting Eq. (\ref{eq:cosh}) into Eq. (\ref{eq:DT}) and integrating over $n$, we find
\begin{equation}
D_{T}\simeq \mathrm{Re}\sum\limits_{k=1}^{\infty}e^{2\pi k i \delta\mu/\hbar \omega_{c}^{*}}\sum\limits_{\pm}\frac{b}{\sqrt{|\epsilon_{n\pm}^{\pm\prime\prime}|T}}e^{-\Delta_{\pm}^{\pm}/T}e^{-k^{2}T/2|\epsilon_{n\pm}^{\pm\prime\prime}|}+\cdots,
\end{equation}
which represents a periodic oscillation with $1/B$. The oscillation frequency $F=\delta\mu m_{d}m_{f}/\hbar^{2}e(m_{d}+m_{f})$ is the same as the unhybridized metal. The oscillation amplitude has the form given in Eq. (9) in the main text. The $e^{-\Delta_{\pm}^{\pm}/T}$ factor comes from the thermal activation and the $e^{-T/2|\epsilon_{n\pm}^{\pm\prime\prime}|}$ factor from the thermal smearing similar to the LK theory.

In the 3D case, $\epsilon_{\pm}^{\pm}$ depends on $n$ and $k_{z}$, so the expansion in Eq. (\ref{eq:expansion}) is replaced by
\begin{equation}
|\epsilon_{\pm}^{\pm}(n,k_{z})-\mu|\simeq \Delta_{\pm}^{\pm}+\frac{1}{2}|\partial_{n}^{2}\epsilon_{\pm}^{\pm}(n_{\mathrm{gap}\pm}^{\pm},0)|(n-n_{\mathrm{gap}\pm}^{\pm})^{2}+\frac{1}{2}|\partial_{k_{z}}^{2}\epsilon_{\pm}^{\pm}(n_{\mathrm{gap}\pm}^{\pm},0)|k_{z}^{2}.
\end{equation}
Similar derivations yield the oscillation amplitude as follows,
\begin{equation}
R_{T}\sim e^{-\Delta/T-bT}.
\end{equation}

\section{Low-energy DOS oscillation in a lattice model}

On a 2D lattice, the two-band model is defined in the momentum space as follows,
\begin{equation}
H=\sum\limits_{k}
\begin{pmatrix}
d_{k}^{\dagger}\\
f_{k}^{\dagger}
\end{pmatrix}
\begin{pmatrix}
-2t(\cos k_{x}a_{0}+\cos k_{y}a_{0})-\mu _{d}& V\vec{s}_{k}\cdot \vec{\sigma}\\
V\vec{s}_{k}\cdot\vec{\sigma}& 2\alpha t(\cos k_{x}a_{0}+\cos k_{y}a_{0})-\mu _{f}
\end{pmatrix}
\begin{pmatrix}
d_{k} &f_{k}
\end{pmatrix},
\end{equation}
in which $\vec{s}_{k}\equiv (\sin k_{x}a_{0},\sin k_{y}a_{0})$. $a_{0}$ is the lattice constant. In the magnetic field, the hopping amplitudes, $t$, $\alpha t$ and $V$ are multiplied by the phase factors induced by the gauge potential, $t_{ij}\mapsto t_{ij}e^{-ieA_{ij}}$. In the Landau gauge, $A_{i,i+\hat{x}}=0$, $A_{i,i+\hat{y}}=Bi_{x}a_{0}^{2}$.

\begin{figure}
\centering
\includegraphics[width=0.6\textwidth]{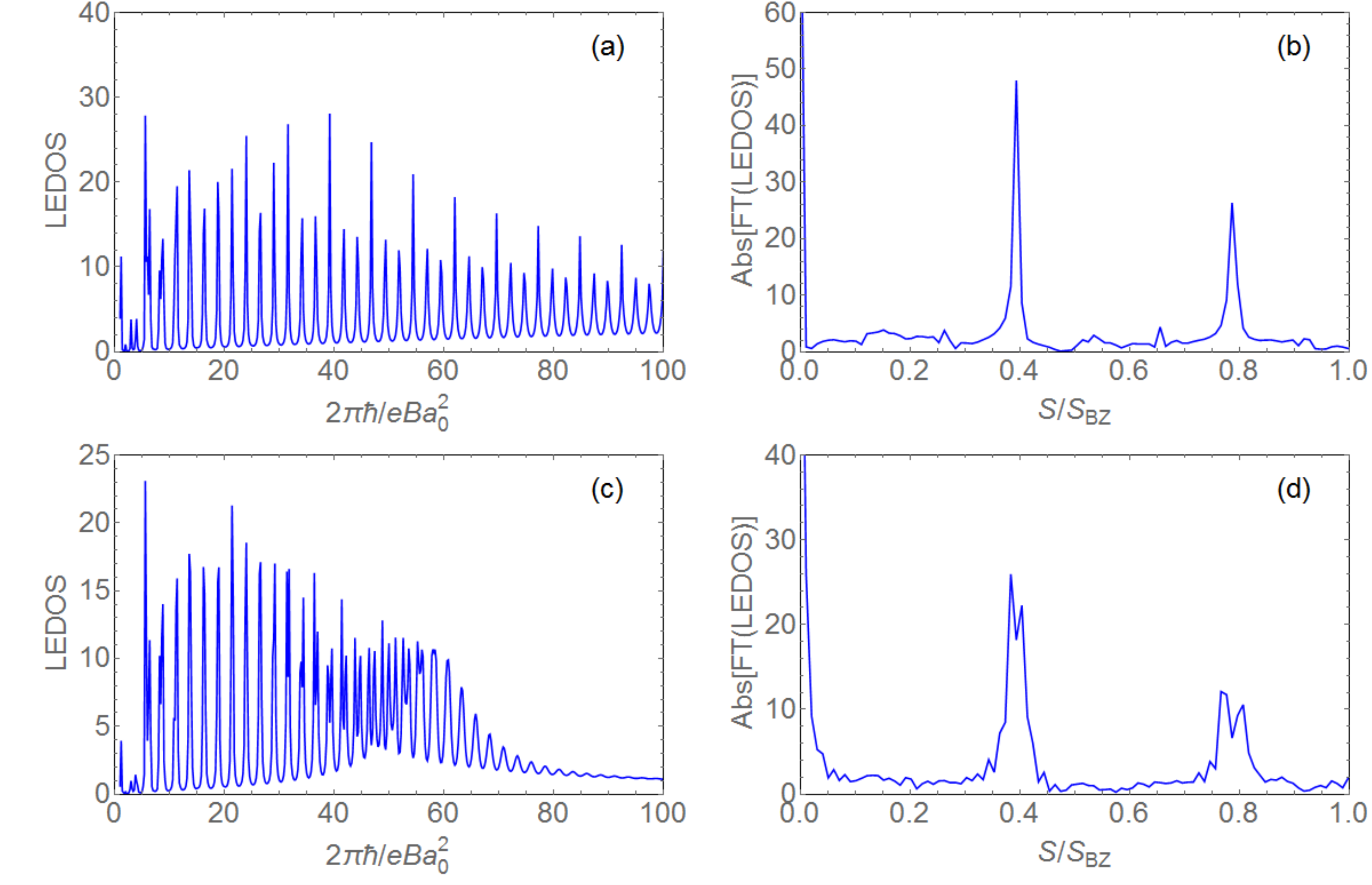}
\caption{LEDOS vs. $1/B$ and its Fourier transformation for (a,b) a metal with $V=0$ and (c,d) a topological insulator with $V=0.015t$ calculated on a lattice with $500\times 500$ sites. The double-peak structure in (d) is an artefact due to the $\pi$ phase change in (c). Band parameters: $\delta \mu\equiv \mu _{d}-\mu _{f}=0.5t$, $\alpha=0.1$. The Lorentzian broadening $\eta=0.001t$.}
\label{fig:lattice}
\end{figure}

For the lattice model, the low-energy density of states on a lattice is defined as
\begin{equation} \label{eq:Dos}
D_{\eta}=\frac{1}{\pi}\mathrm{Im}\mathrm{Tr}\frac{1}{H-i\eta},
\end{equation}
in which the Lorentzian broadening parameter $\eta$ plays the same role as the temperature. This alternative definition can be calculated efficiently with an iterative algorithm on a large lattice \cite{Allais2014, Zhang2014b, Zhang2014c} without explicitly solving the energy spectrum. The results for the metal $V=0$ and the topological insulator with nonzero $V$ are shown in Fig. \ref{fig:lattice}. The LEDOS in both cases show oscillation with the same frequency. For a nonzero hybridization, a relative $\pi$ phase change in the oscillation pattern is found in strong magnetic fields [Fig. \ref{fig:lattice} (c)]. Therefore, the main features of quantum oscillation found in the continuum model are reproduced by the lattice model.

\bibliography{/Dropbox/ResearchNotes/BibTex/library}